\documentclass[final,1p,times]{elsarticle}
\usepackage{color}
\usepackage{amsmath}
\usepackage{subfigure}
\usepackage{lineno,hyperref}

\journal{Physics of the Dark Universe}

\begin{document}
	
	\begin{frontmatter}
		
		\title{Cosmological application of the lens-redshift  probability distribution with improved galaxy-scale gravitational lensing sample}

		\author[address1]{Hui Li}
		
		\author[address2,address3]{Yun Chen\corref{mycorrespondingauthor}}
		\cortext[mycorrespondingauthor]{Corresponding author}
		\ead{chenyun@bao.ac.cn}

		\address[address1]{Gravitational Wave and Cosmology Laboratory, Department of Astronomy, Beijing Normal University, Beijing 100875, China}
		\address[address2]{Key Laboratory for Computational Astrophysics, National Astronomical Observatories, Chinese Academy of Sciences, Beijing 100101, China}
		\address[address3]{College of Astronomy and Space Sciences, University of Chinese Academy of Sciences, Beijing, 100049, China}

\begin{abstract}
We conduct the cosmological analysis by using the lens-redshift distribution test with updated galaxy-scale strong lensing sample, where the considered scenarios involve three typical cosmological models (i.e., $\Lambda$CDM, $\omega$CDM and $\omega_0\omega_a$CDM models) and three typical choices (i.e., non-evolving, power-law and exponential forms) for the velocity-dispersion distribution function (VDF) of lens galaxies. It shows that degeneracies between cosmological and VDF parameters lead to the shifts of estimates on the parameters. The limits on $\Omega_{m0}$ from the lens-redshift distribution are consistent with those from the Pantheon+ Type Ia supernova (SN Ia) sample at 68.3\% confidence level, though the uncertainties on $\Omega_{m0}$ from the former are  about 3 to 8 times larger than those from the latter.  The mean values of $\Omega_{m0}$ shift to the larger values in the power-law VDF case and to the lower values in the exponential VDF case, compared with those obtained in the non-evolving VDF case. In the $\omega$CDM model, the limits on $\omega_0$, i.e. the dark energy equation of state (EoS), are consistent with those from the Pantheon+ sample at 68.3\% confidence level, but the mean values of $\omega_0$ from the former are significantly smaller than those from the latter.  In the $\omega_0\omega_a$CDM model, the uncertainties on $\omega_0$ are dramatically enlarged compared with those obtained in the $\omega$CDM model; moreover, the Markov chains of $\omega_a$, i.e. the time-varying slope of EoS, do not achieve convergence in the three VDF cases. Overall, the lens-redshift distribution test is more effective on constraining $\Omega_{m0}$ than on the dark energy EoS. 
\end{abstract}
		
		
	\end{frontmatter}
	
\section{Introduction}




The galaxy-scale strong gravitational lensing (SGL) has been proved to be a kind of effective cosmological probe, which can be used to study lens properties and to constrain cosmological parameters. In the cosmological implementations of the galaxy-scale SGL, several different quantities have been used as statistical quantities, including the statistics of SGL (see e.g. \cite{Turner_etal_1984,Kochanek_1992}), the velocity dispersions of lenses (i.e., combining the gravitational and dynamical masses of the lens galaxy; see e.g. \cite{Futamase_Yoshida_2001,Biesiada_2006,Chen_etal_2019}), and the SGL time-delay (see e.g. \cite{Refsdal_1964,Treu_Marshall_2016,Bonvin_etal_2017,Birrer_etal_2019}). The continuous increase of the discovered galaxy-scale lensing systems has significantly promoted the development of the corresponding cosmological and astrophysical applications. Especially, the number of confirmed galaxy-scale SGL systems, which own high-resolution imaging data and high-quality spectroscopic redshifts for foreground lenses and background sources, has increased from $\sim$20  in 2006 \cite{Bolton_etal_2006, Koopmans_etal_2006,Treu_etal_2006} to $\sim$160 in 2019\cite{Chen_etal_2019}.
\par
The statistics of SGL mainly include two realization ways\cite{Turner_etal_1984,Ofek_etal_2003}, i.e.,  using the distribution of image separations as a statistic, and using the len-redshift probability distribution as a statistic.
However, the former encounters an obvious bias, which  results from the fact that larger separation lenses are more easily discovered. 
Conversely, in the latter method, the image separation is taken into account as a prior, so almost all the confirmed lenses, regardless of  how they are discovered, can be used in this statistic method. Thus, the latter one is much more popular at present. Recently, a well-selected galaxy-scale SGL sample with 161 systems has been collected by Chen et al. (2019)\cite{Chen_etal_2019}, that is assembled with definite and reasonable criteria to
ensure the selected lenses are early-type galaxies without significant substructures or close massive companions, then the lens galaxies can be well described with the spherically symmetric model.  This sample has been widely used in the cosmological analyses. For example, it has been used to test the  General Relativity\cite{Lian_etal_2022,Liu_etal_2022_GR}, the cosmic distance duality relation\cite{Tang_etal_2023,Qin_etal_2021,Lima_etal_2021,Lyu_etal_2020,Liao_2019} and the invariance of the speed of light\cite{Colaco_etal_2022,Liu_etal_2022_c}, as well as to  estimate the cosmic curvature and other cosmological parameters\cite{Wei_etal_2022,Qi_etal_2021}.
\par
We note that the sample from \cite{Chen_etal_2019} has not been used in the cosmological analysis under the method of using the lens-redshift probability distribution as a statistical quantity. Nevertheless, this method is verified to be effective on constraining both the cosmological model and the velocity-dispersion distribution function of lens galaxies\cite{Turner_etal_1984, Kochanek_1992, Ofek_etal_2003}.  In view of this, we choose to carry out the cosmological analysis via applying the lens-redshift probability distribution test with the updated SGL sample from \cite{Chen_etal_2019}. In this process, we can put constraints on cosmological parameters, as well as explore how the choice for the VDF of lens galaxies affects the estimates of cosmological parameters.
\par
The rest of the paper is organized as follows: in Section \ref{sec:method}, we demonstrate the methodology of using the len-redshift probability distribution to do cosmological analysis, and introduce the cosmological models and the VDF choices for lens galaxies which are taken into account in this work. Then, we conduct the cosmological analysis by using the lens-redshift distribution test with 161 lens systems from \cite{Chen_etal_2019}, and present the main results in Section \ref{sec:analysis}.  In the last section, the main conclusions are summarized.

\section{Methodology and Models}
\label{sec:method} 
\subsection{lens-redshift probability distribution test}
\label{sec:formula} 
The probability of a lensing event,  which is also called as the optical depth for lensing, is given by the product of the number density of lenses and their cross-sections. 
As described in  \cite{Kochanek_1992,Ofek_etal_2003}, the differential optical depth to lensing per unit lens redshift, given the image separation and the source redshift, can be calculated with:
\begin{equation}
\frac{d \tau}{d z_l}=n\left(\theta_E, z_l\right)\left(1+z_l\right)^3 S_{\mathrm{cr}} \left|\frac{c d t}{d z_l}\right|,
\label{eq:tau_z}
\end{equation}
where $n(\theta_E,z_l)$ is the comoving number density of lenses at redshift $z_l$ producing a multiple imaging with Einstein radius $\theta_E$, $S_{cr}$ is the lensing cross-section for multiple imaging, and ${c d t}/{d z_l}$ is the proper distance interval.

For a given lens system, by using the differential optical depth to lensing as the probability density, the relative
probability of finding a lens at redshift $z_l$ with Einstein radius $\theta_E$ for a given source can be derived as
\begin{equation}
\mathcal{P}_i(z_l; \mathbf {p}) =\frac{\int_{z_{l,i}^{\mathrm{min}}}^{z_{l,i}^{\mathrm{max}}}\frac{d \tau}{d z_l} d z_l}{\int_0^{z_{s,i}} \frac{d \tau}{d z_l} d z_l},
\label{eq:relative_prob}
\end{equation}
where $\mathbf {p}$ denotes the model parameter set; $z_{l,i}^{\mathrm{min}} = z_{l,i}-\Delta z/2$, $z_{l,i}^{\mathrm{max}} = z_{l,i}+\Delta z/2$ and the adopted bin width is $\Delta z = 0.1$.
We choose the definition described by Eq. (\ref{eq:relative_prob}) to ensure the value of $\mathcal{P}_i(z_l; \mathbf {p})$ being in the range of 0 to 1. Meanwhile, the normalization term $\int_0^{z_{s,i}} d \tau/d z_l d z_l$ also works as a weight factor, the absence of which may result in biased estimates of the related parameters. 
Obviously, the Hubble constant $H_0$, which is included  in the terms of cosmological distances, is canceled out in the relative probability $\mathcal{P}_i(z_l; \mathbf {p})$.

In this work, it is assumed that the mass profile of  lens galaxy can be represented by the singular isothermal ellipsoid (SIE) model. One can work out the critical radius (i.e., Einstein radius) 
\begin{equation}
\theta_{E} = 4 \pi\left(\frac{\sigma}{c}\right)^2 \frac{D_{\mathrm{ls}}}{D_{\mathrm{s}}}f_E^2,
\label{eq:theta_E}
\end{equation}
where the parameter $f_E$ was initially been brought in to relate the matter velocity dispersion $\sigma$ with the stellar velocity dispersion $\sigma_0$ , i.e., $\sigma=f_E \sigma_0$ \cite{Kochanek_1992}. Nowadays, $f_E$ is extended to be an adaptable parameter to mimic the effects of several systematic errors with $\sqrt{0.8} <f_E<\sqrt{1.2}$\cite{Ofek_etal_2003}. Then, the lensing cross-section can be expressed as
\begin{eqnarray}
	S_{cr}&=&\pi(\theta_E D_l)^2 \nonumber \\
	&=& \pi \left[4 \pi\left(\frac{\sigma}{c}\right)^2 \frac{D_{\mathrm{ls}}}{D_{\mathrm{s}}}f_E^2 \right]^2 D_l^2,
\label{eq:cross_section}
\end{eqnarray}
where $D_l$, $D_s$ and $D_{ls}$ are the angular diameter distances between
the observer and the lens, the lens and the source, and the
observer and the source, respectively.

The distribution of velocity-dispersion of elliptical galaxies is modelled as a modified Schechter function, like
\begin{equation}
\phi(\sigma)\mathrm{d}\sigma=  \phi_*\left(\frac{\sigma}{\sigma_*}\right)^{\alpha}\exp\left[-\left(\frac{\sigma}{\sigma_*}\right)^\beta\right]\frac{\beta}{\Gamma(\alpha/\beta)}\frac{\mathrm{d}\sigma}{\sigma},
\label{eq:Mod_Schechter}
\end{equation}
which was proposed by Sheth et al. (2003)\cite{Sheth_etal_2003} based on the Schechter luminosity function\cite{Schechter_1976},
where $\mathrm{d}n$ is the differential number of elliptical galaxies per unit velocity dispersion per unit comoving volume, 
$\Gamma$ is the gamma function, 
$\phi_*$ is the integrated number density of elliptical galaxies, 
$\sigma_*$ is the characteristic velocity dispersion, 
$\alpha$ is the low-velocity power-law index,
and $\beta$ is the high-velocity exponential cutoff index\cite{Mitchell_etal_2005}. Further, the number density $n(\theta_E,z_l)$ included in Eq.(\ref{eq:tau_z}) can be estimated with the formula:
\begin{equation}
    n(\theta_E,z_l) =\frac{\phi(\sigma)\mathrm{d}\sigma}{\mathrm{d}\theta_E}.
\label{eq:NumberDensity}
\end{equation}

As discussed in \cite{Mitchell_etal_2005}, if $\phi(\sigma)\mathrm{d}\sigma$ of the lens galaxy population is inferred from a combination of the E/S0 galaxy luminosity function and the Faber-Jackson relation, then the ignoring dispersion in the Faber-Jackson relation can lead to a biased estimate of $\phi(\sigma)$ and therefore to the biased and overconfident constraints on the cosmological parameters. Meanwhile, the measured VDF from a large sample of local E/S0 galaxies provides a more reliable method for probing cosmology. The parameters of the VDF measured from the SDSS DR5 sample are provided in \cite{Choi_etal_2007}, i.e., $(\phi_*, \sigma_*, \alpha, \beta) = (8\times10^{-3} \textit{h}^3 \mathrm{Mpc}^{-3}, 161\pm 5\,\mathrm{km}\,\mathrm{s}^{-1}, 2.32\pm 0.10, 2.67\pm 0.07)$. These priors for the local VDF are used in our following analysis.

\subsection{Velocity dispersion function for lens galaxies and Cosmological model}
The early works of applying the lens-redshift  distribution test usually assumed that the VDF of lens galaxies is non-evolutionary \cite{Helbig_Kayser_1996}. Afterwards, in view of the observational indications for evolution in the galaxy luminosity function (e.g. \cite{Cohen_2002}), and in the mass-to-light ratio (e.g. \cite{Keeton_etal_1998, van_Dokkum_etal_2001}),  Ofek et al. (2003)\cite{Ofek_etal_2003} introduced the number and mass evolution of the lens population into the lens-redshift distribution test. However, the VDF adopted in \cite{Ofek_etal_2003} was inferred from the combination of the E/S0 galaxy luminosity function and the Faber-Jackson relation, rather than the measured VDF. 

In this work, we employ the VDF described with Eq. (\ref{eq:Mod_Schechter}), i.e., a modified Schechter function, and adopt the priors for local VDF measured from the SDSS DR5 sample as mentioned in section \ref{sec:formula}.  Furthermore, we analyse and compare the scenarios of non-evolving VDF and evolving VDF parameterized with power-law and exponential forms in the frameworks of three typical cosmological models. 

\subsubsection{Choices for the velocity dispersion function}
To explore the effect of the lens VDF in the lens-redshift distribution test, we consider three typical  choices for the VDF of lens galaxies, i.e., the non-evolving, power-law and exponential forms of VDF. In the non-evolving VDF scenario, the characteristic number density $\phi_\ast$ and the characteristic velocity dispersion $\sigma_\ast$ included in Eq. \eqref{eq:Mod_Schechter} are not varying with redshift, i.e.,
To explore the effect of the lens VDF in the lens-redshift distribution test, we consider three typical  choices for the VDF of lens galaxies, i.e., the non-evolving, power-law and exponential forms of VDF. In the non-evolving VDF scenario, the characteristic number density $\phi_\ast$ and the characteristic velocity dispersion $\sigma_\ast$ included in Eq. \eqref{eq:Mod_Schechter} are not varying with redshift, i.e.,
\begin{equation}
\phi_\ast(z)=\phi_{\ast,0},\quad \sigma_\ast(z)=\sigma_{\ast,0},
\label{eq:non-evolutionary}
\end{equation}
\noindent
where $\phi_{\ast,0}$ and $\sigma_{\ast,0}$ are treated as constants, and the subscript ``0" means the present-day value. In the power-law VDF scenario \cite{Chae_Mao_2003, Matsumoto_Futamase_2008}, $\phi_\ast$ and $\sigma_\ast$ are parameterized as :
\begin{equation}
\phi_\ast(z)=\phi_{\ast,0}(1+z)^{\nu_n},\quad \sigma_\ast(z)=\sigma_{\ast,0}(1+z)^{\nu_{v}},
\label{eq:pl_evolution}
\end{equation}
\noindent
where $\nu_n$ indicates the number density evolutionary slope,  and $\nu_v$ is the velocity dispersion evolutionary slope.
Particularly, note that the case ($\nu_n>0,\nu_{v}<0$) is consistent with the hierarchical model of galaxy formation with bottom-up assembly of structure. 
In the exponential VDF scenario\cite{Ofek_etal_2003, Chae2010, Oguri_etal_2012},  $\phi_\ast$ and $\sigma_\ast$ are expressed as
\begin{equation}
\phi_\ast(z)=\phi_{\ast,0}10^{P_{n}z},\quad \sigma_\ast(z)=\sigma_{\ast,0}10^{U_{v}z},
\label{eq:exp_evolution}
\end{equation}
\noindent
where $P_{n}$ and $U_{v}$ denote the evolutionary slopes of the number density and velocity dispersion, respectively.

\subsubsection{Cosmological models under consideration}
To obtain the expansion history of the universe and the cosmological distances mentioned above, one first needs to have the Friedmann equations for the cosmological models under consideration.
Among the various types of cosmological models, we choose to consider three typical scenarios, e.g., the $\Lambda$CDM model with the dark energy EoS $\omega =p_{\Lambda}/\rho_{\Lambda}=-1$, the $\omega$CDM model with the dark energy EoS $\omega_\mathrm{0} = \mathrm{Constant}$, and the $\omega_0\omega_a$CDM model with a dynamical dark energy EoS parameterized as $\omega(z) = \omega_0 + \omega _az/(1 + z)$ \cite{Chevallier_Polarski2001, Linder_2003}.
The Friedmann equation of the spatially flat $\Lambda$CDM model is
\begin{equation}
E^2(z;\mathbf{p})=\Omega_{m0}(1+z)^3 +1-\Omega_{m0},
\label{eq:Ez_LCDM}
\end{equation}
where $E(z) = H(z)/H_0$ is the reduced Hubble parameter, $H(z)$ is the Hubble parameter and $H_0 = H(z = 0)$ is the Hubble constant .
The Friedmann equation of the spatially flat $\omega$CDM model is
\begin{equation}
E^2(z;\mathbf{p})=\Omega_{m0}(1+z)^3 +(1-\Omega_{m0})(1+z)^{3(1+\omega_\mathrm{0})}.
\label{eq:Ez_wCDM}
\end{equation}
The Friedmann equation of the spatially flat $\omega_0\omega_a$CDM model is
\begin{equation}
E^2(z;\mathbf{p})=\Omega_{m0}(1+z)^3 +(1-\Omega_{m0})(1+z)^{3(1+\omega_0+\omega_a)}\exp{\frac{-3w_az}{1+z}}.
\label{eq:Ez_w0waCDM}
\end{equation}
The last term of Eq.(\ref{eq:tau_z}), i.e. the proper
distance interval ${c d t}/{d z_l}$, can be expressed as, 
\begin{equation}
 \frac{c d t}{d z_l} = -\frac{c}{1+z_l}\frac{1}{H_0E(z_l;\mathbf{p})}.
\label{eq:cdt_dz}
\end{equation}
The angular diameter distances included in  Eq.~(\ref{eq:cross_section}), i.e. $D_l$, $D_s$ and $D_{ls}$, can be formulated as
\begin{equation}
D_l\left(z_l; \mathbf {p}, H_0\right)=\frac{c}{H_0\left(1+z_l\right)} \int_{0}^{z_l} \frac{\mathrm{d} z}{E(z;\mathbf {p})},
\label{eq:D_l}
\end{equation}
\begin{equation}
D_s\left(z_s; \mathbf {p}, H_0\right)=\frac{c}{H_0\left(1+z_s\right)} \int_{0}^{z_s} \frac{\mathrm{d} z}{E(z;\mathbf {p})},
\label{eq:D_s}
\end{equation}
\begin{equation}
D_{ls}\left(z_l, z_{\mathrm{s}} ; \mathbf {p}, H_0\right)=\frac{c}{H_0\left(1+z_{\mathrm{s}}\right)} \int_{z_l}^{z_{\mathrm{s}}} \frac{\mathrm{d} z}{E(z;\mathbf p)}.
\label{eq:D_ls}
\end{equation}

\section{Analysis and Results}
\label{sec:analysis}
We employ the lens-redshift probability distribution test to do the cosmological analysis, where the updated galaxy-scale strong lensing sample compiled by \cite{Chen_etal_2019} is adopted. This sample is  well assembled with definite and reasonable criteria to ensure the selected lenses are early-type galaxies without significant substructures or close massive companions, then the lens galaxies can be well described  with the spherically symmetric model. 
As discussed in section \ref{sec:method}, the scenarios under consideration involve three typical cosmological models (i.e., $\Lambda$CDM, $\omega$CDM and $\omega_0\omega_a$CDM models) and three forms (i.e., non-evolving, power-law and exponential forms) of VDF. 
In our analysis, the total likelihood is assumed to be 
\begin{equation}
\label{eq:LH_total}
\mathcal{L}(\mathbf{p})=\prod_{i=1}^{N_l} \mathcal{P}_i(z_l; \mathbf {p}),
\end{equation}
where $N_l = 161$ is the number of lens systems in the adopted sample, and $\mathcal{P}_i(z_l; \mathbf {p})$ presents the relative probability for each lens system, which is calculated with Eq. (\ref{eq:relative_prob}).
An affine-invariant Markov chain Monte Carlo (MCMC) ensemble sampler (emcee)\cite{ForemanMackey_etal_2012} is employed to generate the posterior probability distributions for the parameters. 

The mean values with 68.3\% confidence limits for the parameters are shown in Table \ref{tab:parameters} for $\Lambda$CDM, $\omega$CDM and $\omega_0\omega_a$CDM models, respectively. The constraints from the lens-redshift distribution test are listed in the upper panels of the tables, and those from the recent Pantheon+ Type Ia supernova sample \cite{Brout_etal_2022} are listed in the lower panels. In the three cosmological models under consideration, the limits on $\Omega_{m0}$ from the lens-redshift distribution test are consistent with those from the recent Pantheon+ Type Ia supernova (SN Ia) sample at 68.3\% confidence level, though the uncertainties on $\Omega_{m0}$ obtained from the former are  about 3 to 8 times larger than those from the latter.  Besides, the mean values of $\Omega_{m0}$ shift to the larger values in the scenario of power-law VDF and to the lower values in the scenario of exponential VDF, compared with those obtained in the scenario of non-evolving VDF. In the $\omega$CDM model, the limits on $\omega_0$, i.e. the dark energy equation of state (EoS), are consistent with those from the Pantheon+ SN Ia sample at 68.3\% confidence level,  but the mean values of $\omega_0$ constrained from the former are significantly smaller than those from the latter.  In the $\omega_0\omega_a$CDM model, the uncertainties on $\omega_0$ are dramatically enlarged  compared with those obtained in the $\omega$CDM model; moreover, the Markov chains of $\omega_a$ do not achieve convergence in all the three VDF forms under consideration, thus the limits on $\omega_a$ at 68.3\% confidence level are not available. In addition, we also use the Bayesian information criterion (BIC) to compare the models. The BIC\cite{{Schwarz_1978}} is defined as
\begin{equation}
\mathrm{BIC} = -2\ln \mathcal{L}_\mathrm{max} + k\ln N,
\end{equation}
where $\mathcal{L}_\mathrm{max}$ is the maximum likelihood, $k$ is the number of the parameters of the considered model, and N is the number of data points used in the fitting. In the framework of BIC, the favourite model is the one with the minimum BIC value. The BIC values for the considered models are also listed in Tables \ref{tab:parameters}. It turns out that the scenario of $\Lambda$CDM model with non-evolving VDF owns the minimum BIC values, which means this scenario is the favourite one.

\begin{figure}[ht!]
		\centering
		\includegraphics[width=\textwidth]{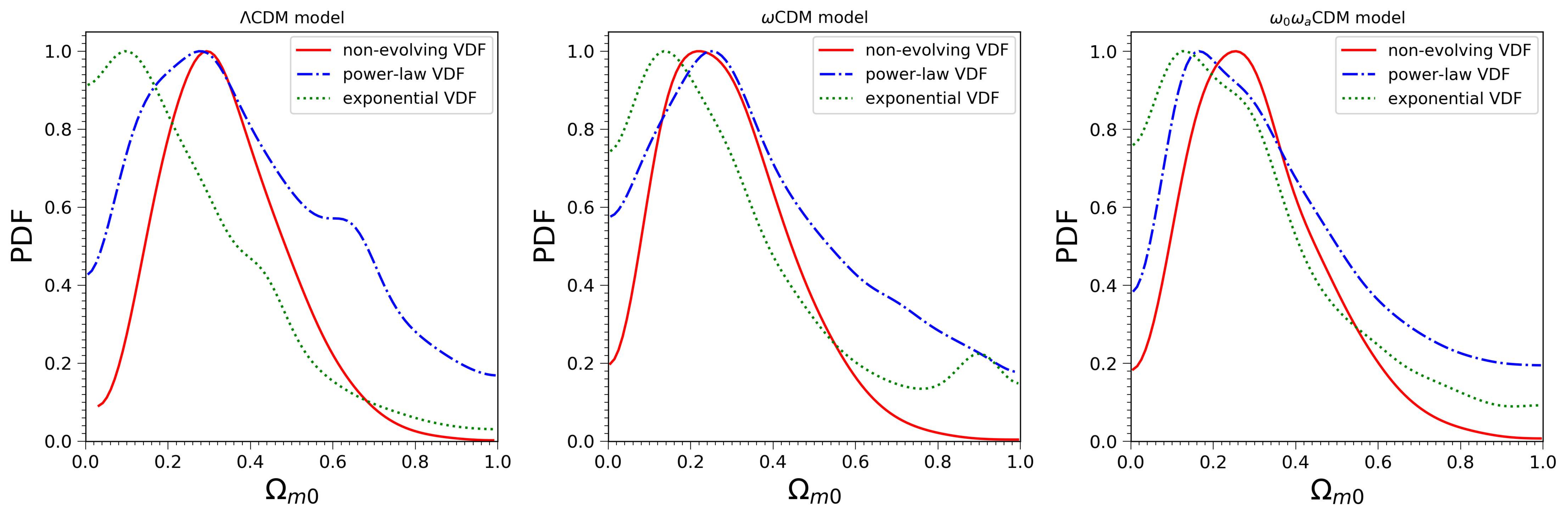}
		 \caption{The 1D marginalized probability distributions of $\Omega_{m0}$ constrained from the lens-redshift distribution test with 161 SGL systems from Chen et al. (2019).}
 \label{fig:Om}
	\end{figure}

\begin{figure}[ht!]
		\centering
		\includegraphics[width=\textwidth]{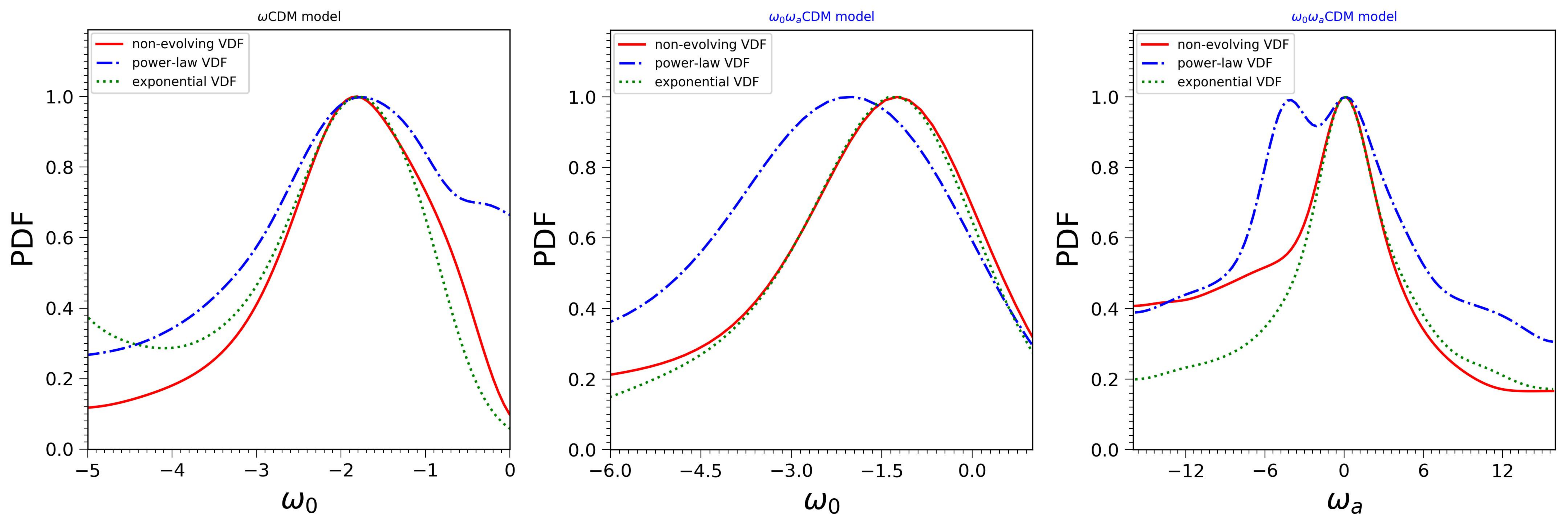}
		 \caption{The 1D  marginalized probability distributions of dark energy EoS parameter(s), constrained from the lens-redshift distribution test for $\omega$CDM model (left panel) and $\omega_0\omega_a$CDM model (middle and right panels), respectively.}
 \label{fig:EoS_w}
	\end{figure}

The one-dimensional (1D) marginalized probability distributions of the involved cosmological parameters are displayed in Figures \ref{fig:Om} and \ref{fig:EoS_w}.   In  Figure \ref{fig:Om}, the 1D distributions of $\Omega_{m0}$ for the three cosmological models show that the maximum likelihood estimates of $\Omega_{m0}$ obtained in the case of exponential VDF are significantly smaller than those obtained in the cases of non-evolving and power-law VDF. In addition, it also displays that the lens-redshift distribution test is more sensitive to the lower limit of $\Omega_{m0}$ than to the upper limit. The 1D distributions of dark energy EoS parameter(s) are demonstrated in Figure \ref{fig:EoS_w} for $\omega$CDM model (left panel) and $\omega_0\omega_a$CDM model (middle and right panels), respectively. In the $\omega$CDM model, the maximum likelihood estimates of $\omega_{0}$ obtained in the three forms of VDF are all much smaller than $-1$, and the dispersion of $\omega_{0}$ obtained in the power-law VDF case is much wider than those obtained in the other two cases. In the $\omega_0\omega_a$CDM  model, the maximum likelihood estimate  of $\omega_{0}$ obtained in the power-law VDF case is much smaller than those obtained in the other two cases; moreover, the Markov chains of $\omega_a$ do not achieve convergence in the range of $\left(-16,16\right)$  for all the three VDF forms under consideration. 

\begin{figure}[htbp]
		\centering
		\includegraphics[width=\textwidth]{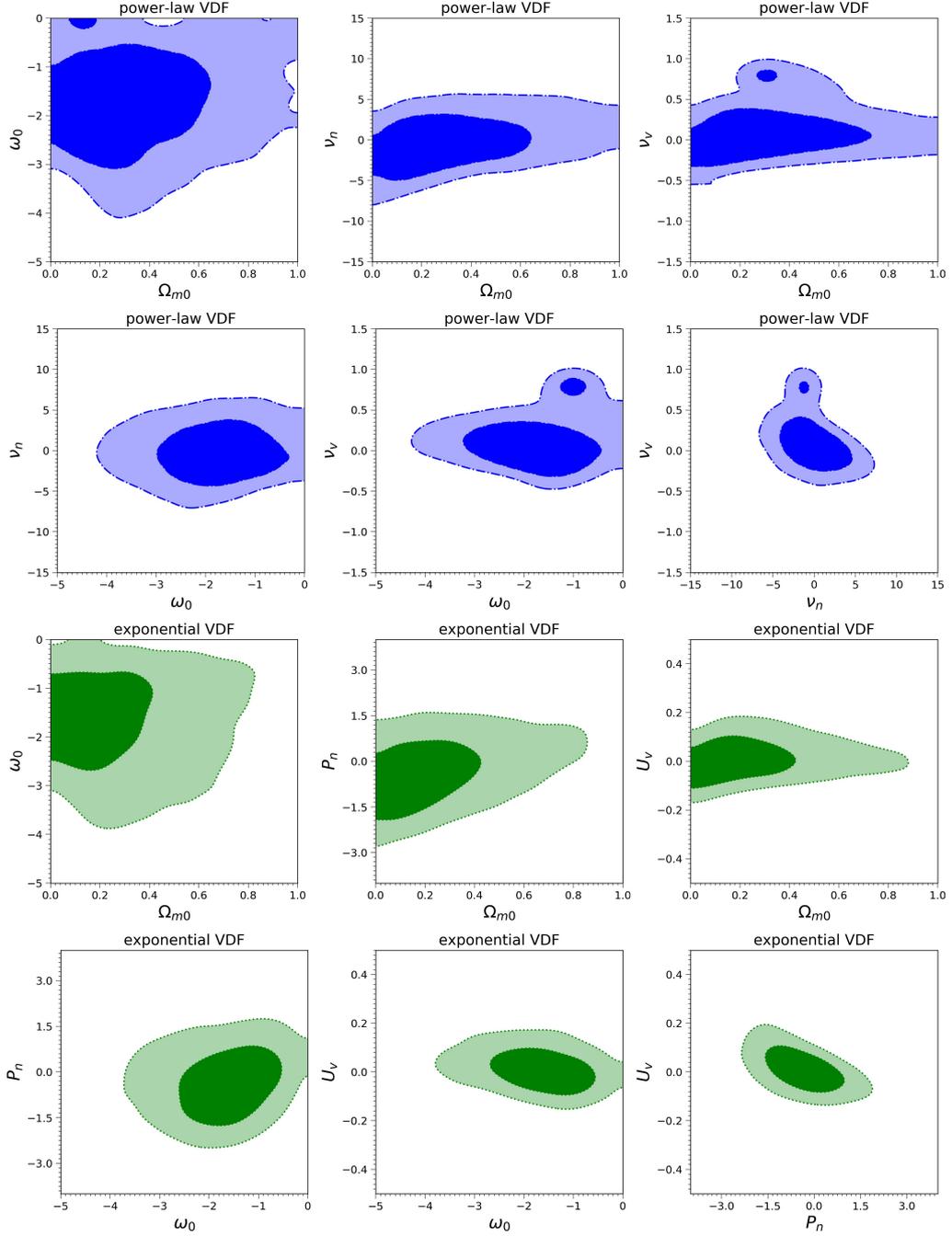}
		\caption{The 2D marginalized  probability distributions of model parameters constrained from the lens-redshift distribution test in the framework of $\omega$CDM model, where the results correspond to the scenarios of power-law VDF (panels in first two rows) and exponential VDF  (panels in last two rows), respectively.}
		\label{fig:wCDM_2D}
	\end{figure}

To explore the degeneracies between the cosmological parameters and the VDF parameters, the  marginalized two-dimensional (2D) posterior distributions of the parameters  in the $\omega$CDM model are shown in Figure \ref{fig:wCDM_2D}. In the scenario of power-law VDF, the visible degeneracies include the positive correlations in the planes of $\Omega_{m0}-\nu_n$ and $\Omega_{m0}-\nu_v$, and the negative correlations in the planes of $\omega_{0}-\nu_v$ and $\nu_n-\nu_v$.
In the scenario of exponential VDF, the visible degeneracies include the positive correlations in the planes of $\Omega_{m0}-P_n$, $\Omega_{m0}-U_v$ and $\omega_{0}-P_n$, and the negative correlations in the planes of $\omega_{0}-U_v$ and $P_n-U_v$.
In general, it turns out that the degeneracies between cosmological parameters and VDF parameters lead  to the shifts of estimates on the cosmological parameters.

\begin{table*}[h!]

    \setlength{\tabcolsep}{3pt} 
    \renewcommand\arraystretch{1.5}
	\centering
	\caption{Observational constrains on the model parameters from the lens-redshift distribution test with 161 lens systems (labeled as ``SGL'') and from the Pantheon+ SNe Ia sample (labeled as ``SNe'' ). All limits and confidence regions quoted here are 68.3 per cent. The reference ``B02'' denotes Brout et al. (2022)\cite{Brout_etal_2022}.}
	\label{tab:parameters}
 \scriptsize
	\begin{tabular}{l|l|llllllll} 
	
	\hline
	 cosmology & data & choice for &$\Omega_\mathrm{m0}$&$\omega_0$&$\omega_a$& parametrs for& & BIC & Reference \\
  &   & VDF & & & &VDF & & & \\
	\hline
    & & non-evolving & $0.321^{+0.166}_{-0.122}$ &...&...& ... & ... & 499.0 &this work\\
    $\Lambda$CDM& SGL & power-law& $0.356^{+0.298}_{-0.205}$ &... &...& $\nu_\mathrm{n} = 0.37^{+2.59}_{-1.78}$ & $\nu_\mathrm{v}=-0.01^{+0.17}_{-0.15}$ & 508.5 & this work\\
    model& & exponential& $0.21^{+0.23}_{- 0.14}$ &...&...& $P_\mathrm{n} = -0.03^{+0.61}_{-0.73}$ & $U_\mathrm{v}=-0.02^{+0.05}_{-0.05}$ & 507.2 & this work\\
   \cline{2-10}
    & SNe & ... & $0.334\pm0.018$ &...&...&... & ...& ...& B02\\
    \hline
    & & non-evolving & $0.272^{+0.169}_{-0.131}$ & $-1.929^{+0.927}_{-1.346}$ &...& ... &...& 500.7 &this work\\
  $\omega$CDM& SGL & power-law& $0.328^{+0.338}_{-0.193}$ &$-1.667^{+1.821}_{-1.966}$ &...& $\nu_\mathrm{n} = 0.69^{+4.23}_{-2.60}$ & $\nu_\mathrm{v}=0.08^{+0.25}_{-0.21}$ & 509.9 & this work\\
    model&& exponential& $0.251^{+0.320}_{-0.158}$ & $-2.327^{+1.046}_{-2.748}$ &...& $P_\mathrm{n} = -0.54^{+0.81}_{-0.76}$ & $U_\mathrm{v}=0.03^{+0.08}_{-0.06}$ & 508.5 & this work\\
  \cline{2-10}
    &SNe & ... & $0.309^{+0.063}_{-0.069}$ &$-0.90\pm0.14$ &...&... & ...& ...& B02\\
    \hline
     && non-evolving & $0.274^{+0.184}_{0.124}$ & $-3.787^{+2.978}_{-7.084}$ & None &... &...& 505.8 &this work\\
    $\omega_0\omega_a$CDM& SGL & power-law& $0.314^{+0.324}_{-0.183}$ &$-3.307^{+2.329}_{-7.162}$ & None & $\nu_\mathrm{n} = -0.43^{+2.77}_{-1.96}$ & $\nu_\mathrm{v}=0.16^{+0.23}_{-0.18}$ & 514.9 & this work\\
    model&& exponential & $0.263^{+0.273}_{-0.148}$ & $-3.701^{+2.779}_{-7.129}$ & None & $P_\mathrm{n} = -0.12^{+1.01}_{-0.89}$ & $U_\mathrm{v}=0.02^{+0.08}_{-0.06}$ & 513.8 & this work\\
   \cline{2-10}
   & SNe&...&$0.403^{+0.054}_{-0.098}$ &$-0.93\pm0.15$ & $-0.1^{+0.9}_{-2.0}$& ...& ...&...& B02\\
    \hline   
	\end{tabular}
\end{table*}
 
\section{Conclusions}
In this work, the lens-redshift probability distribution test is adopted to conduct the cosmological analysis with the updated galaxy-scale strong lensing sample including 161 lenses from \cite{Chen_etal_2019}. To explore the possible degeneracies between the cosmological parameters and the VDF parameters, we have considered the scenarios including three typical cosmological models (i.e., $\Lambda$CDM, $\omega$CDM and $\omega_0\omega_a$CDM models) and three VDF forms (i.e., non-evolving, power-law and exponential forms) for lens galaxies. The MCMC method has been used to compute the posterior probability distributions of the model parameters. The main results can be summarized as follows:
\begin{itemize}
\item The limits on $\Omega_{m0}$ and $\omega_0$ from the lens-redshift distribution test are consistent with those from the Pantheon+ SN Ia sample at 68.3\% confidence level, though the uncertainties on the parameters obtained from the former are much larger than those from the latter. However, the limits on $\omega_a$ at 68.3\% confidence level are not available from the lens-redshift distribution test.
\item  The degeneracies between the cosmological parameters and the VDF parameters are observed, which are the potential factors of leading to the shifts of  estimates on the cosmological parameters. 
\item Though the scenario of non-evolving VDF is preferred in the framework of BIC, there is no robust evidence to exclude the scenarios of evolving VDF.
\item On the whole, the lens-redshift distribution test is  effective on constraining $\Omega_{m0}$, and it behaves not so well on limiting the dark energy EoS. In addition, this test is also reliable on limiting the VDF evolution parameters for lens galaxies.
\end{itemize}

It would be interesting to forecast the power of the lens-redshift distribution test from the upcoming dramatically enlarged SGL sample, which is anticipated in view of the Stage-IV surveys (see e.g. \cite{Oguri_Marshall_2010, Collett2015, Shu_etal_2018} ). 
In addition, a very promising way is to explore the evolution of VDF for lens galaxies with the lens-redshift distribution test,  under the premise of breaking or alleviating the parameter degeneracies with the aid of standard cosmological probes, or adopting the priors of cosmological parameters from the standard cosmological probes.

\section*{Acknowledgments}
HL and YC would like to thank Shuo Cao and Shuaibo Geng for helpful discussions on the velocity dispersion function.
This work has been supported by the National Natural Science Foundation of China (Nos. 11988101 and 12033008), the China Manned Space Project (No. CMS-CSST- 2021-A01), and the K. C. Wong Education Foundation. 


\end{document}